\documentclass[pre,a4paper,amsfonts,amssymb,twocolumn,noshowpacs,noshowkeys,nofootinbib,floats]{revtex4}

\usepackage{amsfonts}
\usepackage{amsmath}
\usepackage{amssymb}
\usepackage{graphicx}

\begin{document}

\title{A nonextensive approach to the dynamics of financial observables}

\author{S\'{\i}lvio M. Duarte Queir\'{o}s}
\email[e-mail address: ]{sdqueiro@cbpf.br}
\affiliation{Centro Brasileiro de Pesquisas F\'{\i}sicas, 150, 22290-180, Rio de Janeiro - RJ, Brazil}
\author{Luis G. Moyano}
\email[e-mail address: ]{moyano@cbpf.br}
\affiliation{Centro Brasileiro de Pesquisas F\'{\i}sicas, 150, 22290-180, Rio de Janeiro - RJ, Brazil}
\author{Jeferson de Souza}
\email[e-mail address: ]{jeferson@cbpf.br}
\affiliation{Centro Brasileiro de Pesquisas F\'{\i}sicas, 150, 22290-180, Rio de Janeiro - RJ, Brazil}
\author{Constantino Tsallis}
\email[e-mail address: ]{tsallis@santafe.edu}
\affiliation{Santa Fe Institute, 1399 Hyde Park Road, Santa Fe - NM, USA}
\affiliation{Centro Brasileiro de Pesquisas F\'{\i}sicas, 150, 22290-180, Rio de Janeiro - RJ, Brazil}

\date{\today}

\begin{abstract}
We present results about financial market observables, specifically 
returns and traded volumes. They are obtained within the current nonextensive statistical mechanical framework
based on the entropy $S_{q}=k\frac{1-\sum\limits_{i=1}^{W} p_{i} ^{q}}{1-q}\, \left( q\in \Re
\right)$ ($S_{1} \equiv S_{BG}=-k\sum\limits_{i=1}^{W}p_{i}\,\ln p_{i}$). 
More precisely, we present stochastic dynamical mechanisms which mimic  
probability density functions empirically observed. These  
mechanisms provide  possible interpretations for the emergence of the entropic 
indices $q$ in the time evolution of the corresponding observables. In addition to this, through multi-fractal analysis 
of return time series, we verify that the dual relation $q_{stat}+q_{sens}=2$  is numerically  satisfied, $q_{stat}$ and $q_{sens}$ being  
associated to the probability density function and to the sensitivity 
to initial conditions respectively. This type of simple relation, whose understanding remains ellusive, has been empirically verified in various other systems.   
\end{abstract}



\maketitle

\section{Introduction}

In recent years statistical mechanics has enlarged its original assignment:
the application of statistics to large systems whose states are governed
by some Hamiltonian functional~\cite{SBG}. Its capability for relating microscopic states of individual constituents of a
system to its macroscopic properties are nowadays used
ubiquitously~\cite{qprops}. Certainly, the most important of these connections still is the determination of thermodynamic properties 
through the correspondence between the entropy concept, originally
introduced by Rudolf Julius Emmanuel Clausius in 1865~\cite{fermi}, and the
number of allowed microscopic states, introduced  by Ludwig Boltzmann around 1877 when
he was studying the approach to equilibrium of an ideal gas~\cite{boltz}.
This connection can be expressed as 
\begin{equation}
S=k\,\ln W,  \label{boltz-prin}
\end{equation}%
where $k$ is a positive constant, and $W$ is the number of microstates
compatible with the macroscopic state of an isolated system. This equation, known as {\it Boltzmann
principle}, is one of the cornerstones of standard statistical mechanics.

When the system is not isolated, but instead in contact to some large 
reservoir, it is possible to extend Eq. (\ref{boltz-prin}), under some
assumptions, and obtain the {\it Boltzmann-Gibbs entropy} 
\begin{equation}
S_{BG}=-k\sum\limits_{i=1}^{W}p_{i}\,\ln p_{i} \,,
\end{equation}
where $p_{i}$ is the probability of the microscopic configuration $i$~\cite%
{SBG}. The Boltzmann principle should  be derivable from
microscopic dynamics, since it refers to microscopic states, but the
implementation of such calculation has not been yet achieved. So,
Boltzmann-Gibbs (BG) statistical mechanics is still based on hypothesis such as the molecular chaos~\cite{boltz} 
and ergodicity~\cite{khinchin}. In spite of the lack of an actual fundamental derivation, 
BG statistics has been undoubtedly successful in the treatment of systems in which {\it short} spatio/temporal
interactions dominate. For such cases, ergodicity and (quasi-) independence are
favoured and Khinchin's approach to $S_{BG}$ is valid~\cite{khinchin}.
Therefore, it is entirely feasible that other physical entropies, in addition to the BG one, can be defined in order to properly treat anomalous
systems, for which the simplifying hypothesis of ergodicity and/or independence are not fulfilled. 
Examples are: metastable states in long-range
interacting Hamiltonian dynamics, metaequilibrium states in small systems
(i.e., systems whose number of particles is much smaller than Avogrado's
number), glassy systems, some types of dissipative dynamics, and other systems that in
some way violate ergodicity. This includes systems with non-Markovian memory (i.e., long-range memory), like it
seems to be the case of financial ones. Generically speaking, systems that may
have a multi-fractal, scale-free or hierarchical structure in the occupancy
of their phase space.

Inspired by this kind of systems it was proposed in $1988$ the entropy~\cite%
{ct} 
\begin{equation}  \label{Sq}
S_{q}=k\frac{1-\sum\limits_{i=1}^{W} p_{i} ^{q}}{q-1}\qquad \left( q\in \Re
\right) ,
\end{equation}
which generalises $S_{BG}$ ($\lim_{q\rightarrow 1}S_{q}=S_{BG}$),
as the basis of a possible generalisation of BG statistical
mechanics\cite{further,3-gen}. The value of the \textit{entropic index }$q$ for a specific system 
is to be determined \textit{a priori} from microscopic dynamics. Just like $%
S_{BG}$, $S_{q}$ is \textit{nonnegative}, \textit{concave}, \textit{%
experimentally robust} (or \textit{Lesche-stable}~\cite{props}) ($\forall {%
q>0}$), and leads to a \textit{finite entropy production per unit time}~\cite%
{qprops,latorabaranger99}. Moreover, it has been recently shown~\cite%
{additive} that it is also \textit{extensive}, i.e., 
\begin{equation}
S_{q}\left( A_{1}+A_{2}+\ldots +A_{N}\right)
\simeq \sum\limits_{i=1}^{N}S_{q}\left( A_{i}\right),
\end{equation}
for special kinds of \textit{correlated} systems, more precisely when the 
phase-space is occupied in a scale-invariant form.  By being extensive, for an appropriate value of $q$,  $S_q$  complies with
Clausius' concept on macroscopic entropy, and with thermodynamics.

Since its proposal, entropy~(\ref{Sq}) has been the source of several
results in both fundamental and applied physics, as well as in other
scientific areas such as biology, chemistry, economics, geophysics and
medicine~\cite{applications}. Herein, we both review and present some
new results concerning applications to the dynamics of financial market
observables, namely the price fluctuations and traded volumes. Specifically,
we will introduce stochastic dynamical mechanisms which are able to
reproduce some features of quantities such as the probability density
functions (PDFs) and the Kramer-Moyal moments. Moreover, we will present some
results concerning the return multi-fractal structure, and its relation to sensitivity to initial conditions. 

Our dynamical proposals will be faced to
empirical analysis of 1 minute returns and traded volumes of the 30 companies
that were used to compose the Dow Jones Industrial Average (DJ30) between the 
$1^{st}$ July and the $31^{st}$ December $2004$. In order to eliminate specious
behaviours we have removed the well-known intra-day pattern following a
standard procedure~\cite{pattern}. After that, the return values were
subtracted from its average value and expressed in standard deviation units,
whereas the traded volumes are expressed in mean traded volume units.

\section{Variational principle using the entropy $S_{q}$}

Before dealing with specific financial problems, let us analyse the probability
density function which emerges when the variational principle is applied to $%
S_{q}$ \cite{3-gen}.

Let us consider its continuous version, i.e., 
\begin{equation}
S_{q}=k\frac{1-\int \left[ p\left( x\right) \right] ^{q}\ dx}{1-q}.
\label{sq-cont}
\end{equation}%
The natural constraints in the maximisation of (\ref{sq-cont}) are 
\begin{equation}
\int p\left( x\right) \ dx=1\,,
\end{equation}
corresponding to normalisation, and 
\begin{equation}
\int x\frac{\ \left[ p\left( x\right) \right] ^{q}}{\int \left[ p\left(
x\right) \right] ^{q}dx}\ dx\equiv \left\langle x\right\rangle _{q}=\bar{\mu}%
_{q}\,,  \label{mean}
\end{equation}%
\begin{equation}
\int \left( x-\bar{\mu}_{q}\right) ^{2}\frac{\ \left[ p\left( x\right) %
\right] ^{q}}{\int \left[ p\left( x\right) \right] ^{q}dx}\ dx\equiv
\left\langle \left( x-\bar{\mu}_{q}\right) ^{2}\right\rangle _{q}=\bar{\sigma%
}_{q}^{2}\,,  \label{variance}
\end{equation}%
corresponding to the \textit{generalised} mean and variance of $x$,
respectively~\cite{3-gen}. 

From the variational problem using (\ref{sq-cont}) under the above
constraints, we obtain 
\begin{equation}  \label{pq-1}
p\left( x\right) =\mathcal{A}_{q}\left[ 1+\left( q-1\right) \mathcal{B}
_{q}\left( x-\bar{\mu}_{q}\right) ^{2}\right] ^{\frac{1}{1-q}},\qquad \left(
q<3\right),
\end{equation}
where, 
\begin{equation}
\mathcal{A}_{q}=\left\{ 
\begin{array}{ccc}
\frac{\Gamma \left[ \frac{5-3q}{2-2q}\right] }{\Gamma \left[ \frac{2-q}{1-q} %
\right] }\sqrt{\frac{1-q}{\pi }\mathcal{B}_{q}} & \Leftarrow & q<1 \\[3mm] 
\frac{\Gamma \left[ \frac{1}{q-1}\right] }{\Gamma \left[ \frac{3-q}{2q-2} %
\right] }\sqrt{\frac{q-1}{\pi }\mathcal{B}_{q}} & \Leftarrow & q>1%
\end{array}
\right. ,
\end{equation}
and
\begin{equation}
\mathcal{B}_{q}=\left[ \left( 3-q\right) \,\bar{\sigma}_{q}^{2}\right] ^{-1}.
\end{equation}
Standard and generalised variances, $\bar{\sigma}%
^{2}$ and $\bar{\sigma}_{q}^{2}$ respectively,  are related by 
\begin{equation}
\bar{\sigma}_{q}^{2}=\bar{\sigma}^{2} \frac{5-3q}{3-q} \,.
\end{equation}

Defining the $q$-\textit{exponential} function as 
\begin{equation}
e_{q}^{x} \equiv \left[ 1+\left( 1-q\right) \,x\right] ^{ \frac{1}{1-q}%
}\qquad \left( e_{1}^x \equiv e^{x}\right) ,
\end{equation}
($e_q^x=0$ if $1+(1-q)x \le0$) we can rewrite PDF~(\ref{pq-1}) as 
\begin{equation}  \label{pq}
p\left( x\right) =\mathcal{A}_{q}\,e_{q}^{-\mathcal{B}_{q}\left( x-\bar{\mu}%
_{q}\right) ^{2}},
\end{equation}
hereafter referred to as $q$-\emph{Gaussian}.

For $q=\frac{3+m}{1+m}$, the $q$-Gaussian form recovers the Student's $t$%
-distribution with $m$ degrees of freedom ($m=1,2,3,\ldots $) with finite
moment up to order $m^{th}$. So, for $q>1,$ PDF~(\ref{pq}) presents an
asymptotic \textit{power-law} behaviour. On the other hand, if $q=\frac{n-4}{%
n-2}$ with $n=3,4,5,\ldots $, $p\left( x\right) $ recovers the $r$%
-distribution with $n$ degrees of freedom. Consistently, for $q<1$, $%
p\left(x\right)$ has a \textit{compact support} which is defined by the
condition $\left\vert x-\bar{\mu}_{q}\right\vert \leq \sqrt{\frac{3-q}{1-q}\,%
\bar{\sigma}_{q}^{2}}$.

\section{Application to macroscopic observables}

\subsection{Model for price changes}

The Gaussian distribution, recovered in the limit $q\rightarrow 1$ of
expression (\ref{pq}), can be derived from various standpoints. Besides the
variational principle, it has been derived, through dynamical arguments, by
L. Bachelier in his 1900 work on price changes in Paris stock market~\cite%
{bachelier}, and also by A. Einstein in his 1905 article on Brownian motion~%
\cite{einstein}. In particular, starting from a Langevin dynamics, we are
able to write the corresponding Fokker-Planck equation and, from it, to
obtain as solution the Gaussian distribution. Analogously, it is also
possible, from certain classes of stochastic differential equations and
their associated Fokker-Planck equations, to obtain the distribution given
by Eq.~(\ref{pq}).

In this section, we will discuss a dynamical mechanism for returns, $r$, which is
based on a Langevin-like equation that leads to a PDF ($q$-Gaussian) with asymptotic
power-law behaviour ~\cite{borland-pre,smdq-quantf}. This
equation is expressed as 
\begin{equation}
dr=-k\,r\,dt+\sqrt{\theta \,\left[ p\left( r,t\right) \right] ^{\left(
1-q\right) }}\,dW_{t} \qquad \left( q\geq 1\right) ,  \label{langevin}
\end{equation}
(in It\^{o} convention) where $W_t$ is a regular Wiener process and $p(r,t)$ is the intantaneous return PDF. 
In a return context the deterministic term of eq. (\ref{langevin}) intends
to represent internal mechanisms which tend to keep the market in some
average return or, in a analogous interpretation, can be related to the eternal competition between speculative
price and the actual worth of a company. In our case, we use the simplest
approach and write it as a restoring force, with a constant $k$, similar to
the viscous force in the regular Langevin equation. In regard to the
stochastic term, it aims to reproduce the microscopic response of the system
to the return: $\theta $ is the volatility constant (intimately associated 
to the variance of $p(r,t)$) and $q$, the nonextensive index, 
reflects the magnitude of that response.
Since the largest unstabilities in the market are introduced
by the most unexpected return values, it is plausible that the stochastic
term in Eq.~(\ref{langevin}) can have such inverse dependence on the PDF $p(r,t)$. Furthermore, Eq.~(\ref{langevin}) presents a 
dynamical
multiplicative noise structure given by,
\begin{equation}
r\left( t\right) =\int_{-\infty }^{t}e^{-k\left( t-t^{\prime }\right) }\sqrt{%
\theta \left[ p\left( r,t\right) \right] ^{\left( 1-q\right) }}%
\;dW_{t^{\prime }} \,,  \label{langevin-solution}
\end{equation}
where we have assumed $r\left( -\infty \right) =0$.

The associated Fokker-Planck equation to Eq.~(\ref{langevin}) is given by  
\begin{equation}
\frac{\partial p(r,t)}{\partial t}=\frac{\partial }{\partial r}\left[
k\,r\,p(r,t)\right] +\frac{1}{2}\frac{\partial ^{2}}{\partial r^{2}}\left\{
\theta \,\left[ p\left( r,t\right) \right] ^{\left( 2-q\right) }\right\} ,
\label{fokker-planck}
\end{equation}
and the long-term probability density function is\cite{borland-pre,risken,plastinos}, 
\begin{equation}
p\left( r\right) =\frac{1}{Z}\left[ 1-\left( 1-q\right) \frac{k\,r^{2}}{%
\left( 2-q\right) \,Z^{q-1}\,\theta }\right] ^{\frac{1}{1-q}} \,.
\label{prob-fokker}
\end{equation}
One of the most interesting features of eq.~(\ref{langevin}) is its aptitude
to reproduce the celebrated U-shape of the $2^{nd}$ (i.e., $n=2$) Kramers-Moyal moment
\begin{equation}
M_{n}\left( r,t,\tau \right) =\int \left( r^{\prime }-r\right) ^{n}\,P\left(
r^{\prime },t+\tau |r,t\right) \,dr^{\prime }\approx \tau \,\theta \,\left[
p\left( r,t \right) \right] ^{\left( 1-q\right) } \,.  \label{m2-eq}
\end{equation}
It is this fact which allowed the establishment of analogies (currently used in financial mimicry) between financial markets
dynamics and fluid turbulence ~\cite{gash}.

It is noteworthy that eq. (\ref{langevin-solution}) is statistically
equivalent to \cite{risken,celia-ct} 
\begin{equation}
dr=-k\,r\,dt+A_{q}\left( t\right) \,dW_{t}+\left( q-1\right) B_{q}\left(
r,t\right) \,dW_{t}^{\prime } \,,  \label{adi-multi}
\end{equation}%
\textit{i.e.}, a stochastic differential equation with {\it independent} additive
and multiplicative noises. If eq. (\ref{langevin}) allows an immediate
heuristic relation between $q$ and the response of the system to its own
dynamics, eq. (\ref{adi-multi}) permits a straighforward dynamical relation
between $q$ and the magnitude of multiplicative noise in such a way that, for $%
q=1$, the Langevin equation is recovered as well as the Gaussian distribution.

\begin{figure}[tbp]   
\begin{center}
\includegraphics[width=0.8\columnwidth,angle=0]{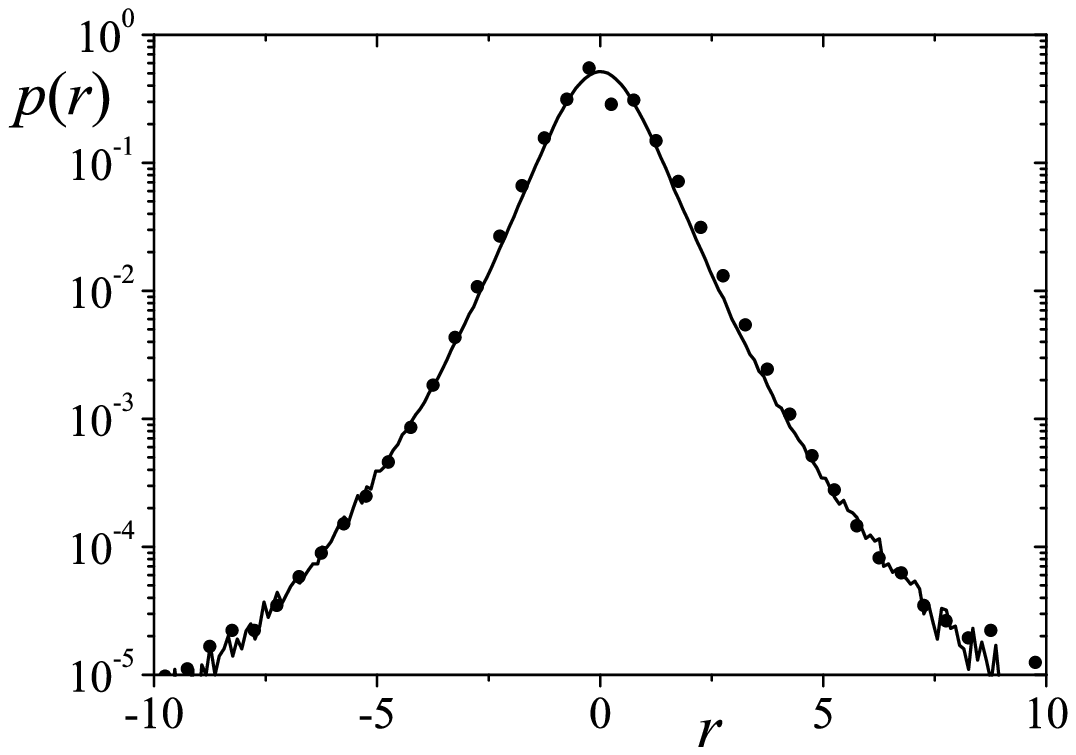}
\includegraphics[width=0.8\columnwidth,angle=0]{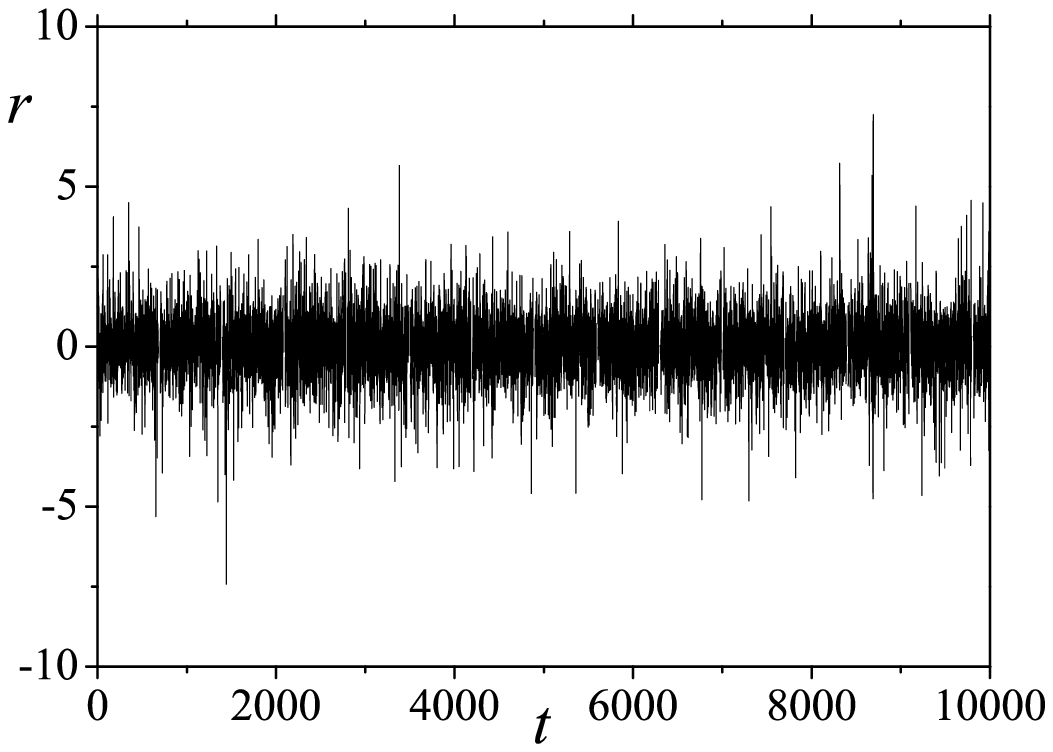}
\includegraphics[width=0.8\columnwidth,angle=0]{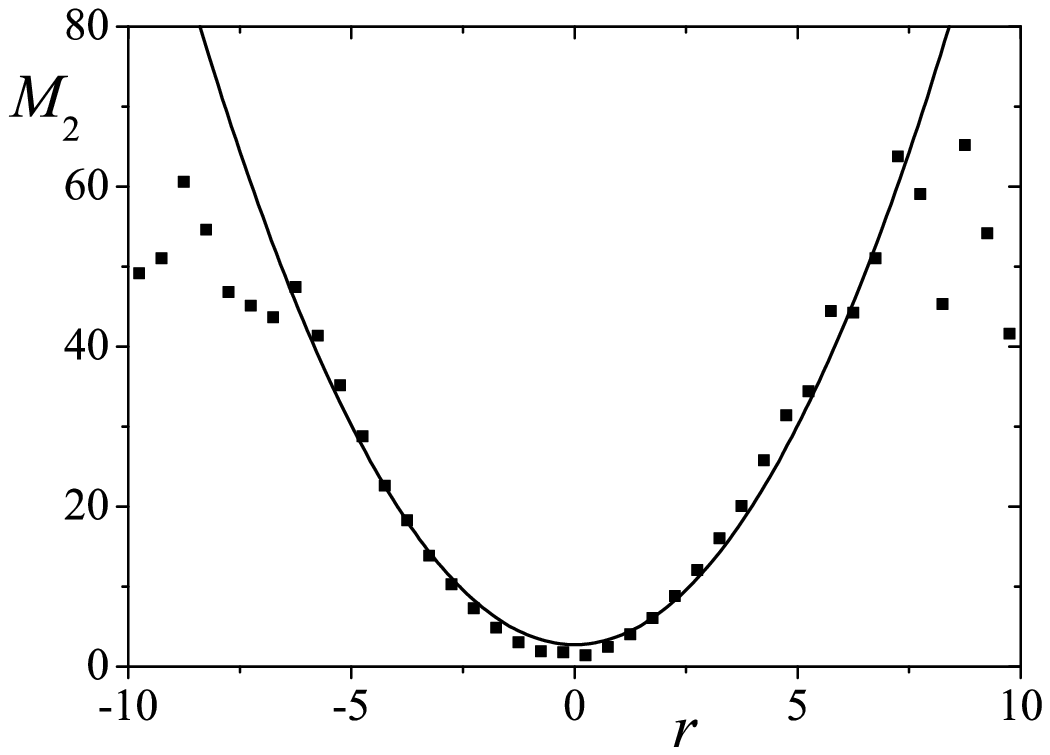}
\end{center}
\caption{
Upper panel: Probability density function \textit{vs.} $r$. Symbols correspond to an average over the
30 equities used to built DJ30 and the line represents the PDF obtained from a time series generated 
by eq.~(\ref{langevin-solution}) which is
presented on middle panel. Lower panel: $2^{nd}$ Kramers-Moyal moment $%
M_{2}\approx \tau \,\theta \,\left[ p\left( r\right) \right] ^{\left(
1-q\right) }=\tau \frac{k}{2-q}\left[ \left( 5-3\,q\right) \sigma
^{2}+\left( q-1\right) r^{2}\right] $ from which $k$ parameter is obtained and where 
the stationary hypothesis is assumed ($t_{0} = - \infty \ll -k^{-1} \ll 0$).
Parameter values: $\tau =1\min $, $k=2.40\pm 0.04$, $\sigma =0.930\pm 0.08$
and $q=1.31\pm 0.02$.
} \label{fig-1}
\end{figure}

In Fig.~(\ref{fig-1}) we present the typical PDF for the 1 minute returns of
a company constituent of the Dow Jones Industrial Average 30 (upper panel)
presenting $q=1.31\pm 0.02$, a time series generated by eq.~(\ref{langevin})
 (middle panel), and the U-shaped $2^{nd}$ Kramers-Moyal moment for our data
(lower panel). As it can be seen the accordance using the simplest approach
is already quite nice. Upgrades of this model can be obtained by taking into account the
risk-aversion effects, which induce asymmetry on the PDF, and correlations on
the volatility in a way which differs from others previously proposed. The
formulation presented herein has also the advantage of being aplicable to
systems which are not in a stationary state since the time-dependent
solutions of the Fokker-Planck equation are of the $q$-Gaussian type as well.

\subsection{Model for traded volumes}

Changes in the price of a certain equity are naturally dependent on
transactions of that equity and thus on its traded volume, $v$. Previous studies
proved the asymptotic power-law behaviour of traded volume PDF \cite{gopi}, 
later extended for all values of $v$ \cite{obt}. In this case it was shown
that the traded volume PDF is very well described by the following ansatz
distribution 
\begin{equation}
P\left( v\right) =\frac{1}{Z}\left( \frac{v}{\varphi }\right) ^{\rho }\exp
_{q}\left( -\frac{v}{\varphi }\right) ,  \label{prob-obt}
\end{equation}%
where $v$ represents the traded volume expressed in its mean value unit $%
\left\langle V\right\rangle $, i.e., $v=V/\langle V\rangle $, $\rho $ and $%
\varphi $ are parameters, and  $Z=\int_{0}^{\infty }\left( 
\frac{v}{\varphi }\right) ^{\rho }\exp _{q}\left( -\frac{v}{\varphi }\right)
\,dv$.

\begin{figure}[tbp]
\begin{center}
\includegraphics[width=0.8\columnwidth,angle=0]{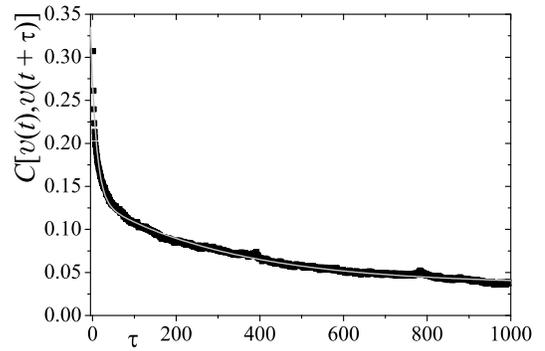}
\end{center}
\caption{
Symbols represent the average correlation function for the 30 time series analysed 
and the line represents a double exponential fit with characteristic times of 
$T_{1}=13$ and $T_{2} = 332$ yielding a ratio about $25$ between the
two time scales Eq.~(\ref{correlation}) ($R^2 = 0.991$, $\chi ^2 = 9 \times 10^{-6}$, 
and time is expressed in minutes).
} \label{fig-2}
\end{figure}

The probability density function (\ref{prob-obt}) was recently obtained from
a mesoscopic dynamical scenario \cite{smdq-vol} based in the following multiplicative noise
stochastic differential equation 
\begin{equation}
dv=-\gamma (v-\frac{\omega }{\alpha })\,dt+\sqrt{2\frac{\,\gamma \,}{\alpha }%
}v\,dW_{t} \,,  \label{feller}
\end{equation}%
where $W_{t}$ is a regular Wiener process following a normal distribution, 
and $v\geq 0$. The right-hand side terms of~eq. (\ref{feller}) represent  
inherent mechanisms of the system in order to keep  $v$ close to some
\textquotedblleft normal\textquotedblright\ value, $\omega /\alpha $, and to
mimic microscopic effects on the evolution of $v$, like a
multiplicative noise commonly used in intermittent processes. This dynamics, 
and the corresponding Fokker-Planck equation~\cite{risken}, lead to the following 
inverted Gamma stationary distribution:
\begin{equation}
f\left( v\right) =\frac{\,\,1}{\omega \,\Gamma \left[ \alpha +1\right] }%
\left( \frac{v}{\omega }\right) ^{-\alpha -2}\,\exp \left[ -\frac{\,\omega }{%
v}\right] .  \label{f-v}
\end{equation}%
Consider now, that instead of being a constant, $\omega $ is a time dependent
quantity which evolves on a time scale $T$ larger than the time scale of
order $\gamma ^{-1}$ required by eq.~(\ref{feller}) to reach stationarity 
\cite{beck-cohen,epl-volume}. This time dependence is, in the present model,
associated to changes in the volume of activity (number of traders that
performed transactions) and empirically justified through the analysis of
the self-correlation function for returns. In Fig.~\ref{fig-2} we have verified
that the correlation function is very well described by
\begin{equation}
C\left[ v\left( t\right) ,v\left( t+\tau \right) \right] =C_{1}\,e^{-\tau
/T_{1}}+C_{2}\,e^{-\tau /T_{2}}
\label{correlation}
\end{equation}
with $T_{2}=332\gg T_{1}=13$. In other words, there is first a fast decay
of $C\left[ v\left( t\right) ,v\left( t+\tau \right) \right] $, related to
local equilibrium, and then a much slower decay for larger $\tau $. This constitutes a necessary condition for the application 
of a superstatistical model~\cite{beck-cohen}.

If we assume that $\omega $ follows a Gamma PDF, i.e., 
\begin{equation}
P\left( \omega \right) =\frac{1}{\lambda \Gamma \left[ \delta \right] }%
\left( \frac{\omega }{\lambda }\right) ^{\delta -1}\exp \left[ -\frac{\omega 
}{\lambda \,}\right] \,, \label{p-omega}
\end{equation}%
then, the long-term distribution of $v$ will be given by $p\left( v\right)
=\int f\left( v\right) \,P\left( \omega \right) \,d\omega $.  This results in  
\begin{equation}
p\left( v\right) =\frac{1}{Z}\left( \frac{v}{\theta }\right) ^{-\alpha
-2}\exp _{q}\left[ -\frac{\theta }{v}\right] \,, \label{p-v}
\end{equation}%
where $\lambda =\theta \left( q-1\right) $, $\delta =\frac{1}{q-1}-\alpha -1$%
. Bearing in mind that, for $q>1$,%
\begin{equation}
x^{a}e_{q}^{-\frac{x}{b}}=\left[ \frac{b}{q-1}\right] ^{\frac{1}{q-1}}x^{a-%
\frac{1}{q-1}}\,e_{q}^{-\frac{b/(q-1)^{2}}{x}},  \label{gamma-transform}
\end{equation}%
we can redefine our parameters and obtain the $q$-Gamma PDF (\ref{prob-obt}).

\begin{figure}[tbp]
\begin{center}
\includegraphics[width=0.8\columnwidth,angle=0]{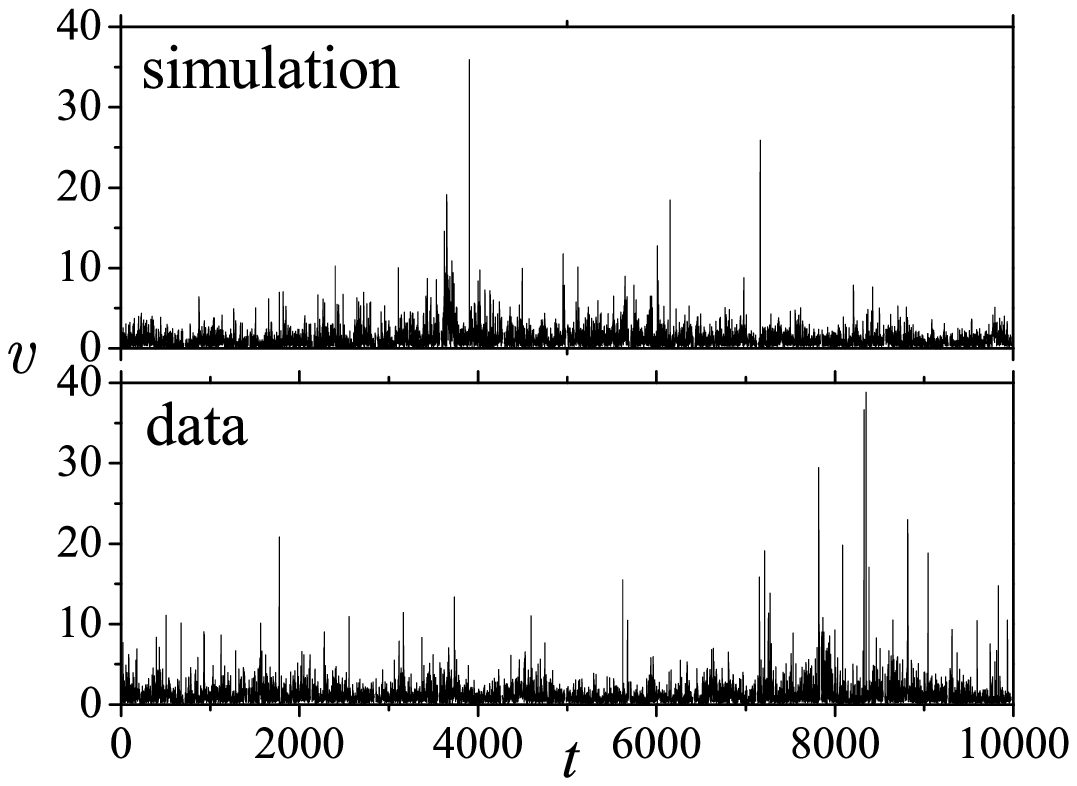}
\includegraphics[width=0.8\columnwidth,angle=0]{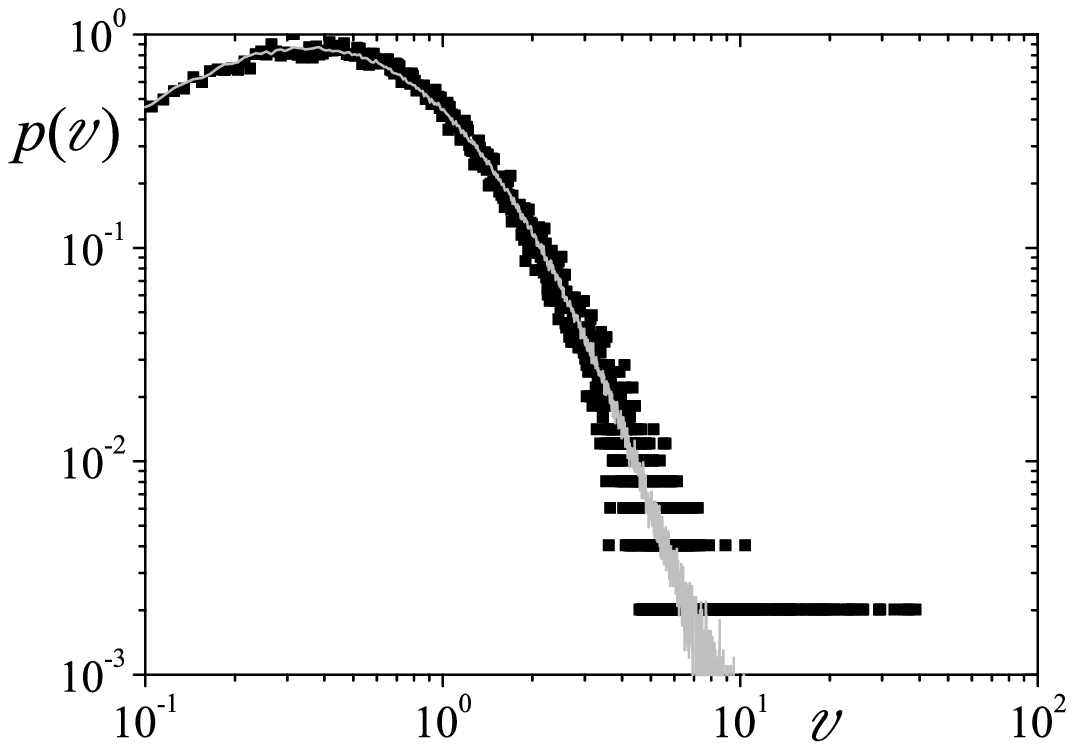}
\end{center}
\caption{Upper panel: Excerpt of the time series generated by our dynamical mechanism (simulation)
 to replicate 1 min traded volume of Citigroup stocks at NYSE (data). Lower panel: 1 min traded volume of Citigroup stocks   
probability density function \textit{vs.} traded volume. Symbols are for 
data, and solid line for the replica. Parameter values: $\theta = 0.212 \pm 0.003$, 
$\rho = 1.35 \pm 0.02$, and $q = 1.15 \pm 0.02$ ($\chi ^2 = 3.6 \times 10^{-4}$, $R^2=0.994$).}
\label{fig-3}
\end{figure}

In Fig.~\ref{fig-3}  we present a comparation between the traded volume of
Citigroup (2004 world's number one company \cite{forbes}) stocks, as well as a replica
of that time series obtained using this dynamical proposal. As it can be
easily verified, the agreement is remarkable.

\section{The nonextensive $q$-triplet and financial observables}

Systems characterised by Boltzmann-Gibbs statistical mechanics present the
following characteristics: (i) Their PDF for energies is proportional to an {\it exponential}
function in the presence of a thermostat; (ii) They have strong sensitivity to the initial conditions, i.e., this quantity increases {\it exponentially} with time (currently referred to as {\it strong chaos}), being 
characterised by a positive maximum Lyapunov exponent; (iii) They typically present, for basic macroscopic quantities, an {\it exponential decay} with some relaxation time. 
In other words, these three behaviours exhibit exponential
functions (i.e., $q=1$). Analogously, it was recently conjectured \cite{ct-vila}
that, for systems which can be studied within nonextensive statistical
mechanics, the energy probability density function (associated to
stationarity or (meta)
equilibrium), the sensitivity to the initial conditions, and the relaxation would be described
by three entropic indices $q_{stat}$, $q_{sens}$, and $q_{rel}$, referred to as the $q$%
\textit{-triplet}. The first physical corroboration of such scenario has
been made from the analysis of two sets of daily averages of the magnetic
field strength observed by Voyager 1 in the solar wind \cite{solar-wind}.
Others systems are currently on study (e.g., \cite{ernesto-bruce}). Of course, if the system is non Hamiltonian, it has no energy distribution, hence $q_{stat}$ cannot defined in this manner. We may however estimate it through a stationary state generalised Gaussian (which would generalise the Maxwellian distribution of velocities for a BG system in thermal equilibrium). In contrast, the other two indices, $q_{sens}$ and $q_{rel}$, remain defined in the usual way. 

\begin{figure}[tbp]
\begin{center}
\includegraphics[width=0.8\columnwidth,angle=0]{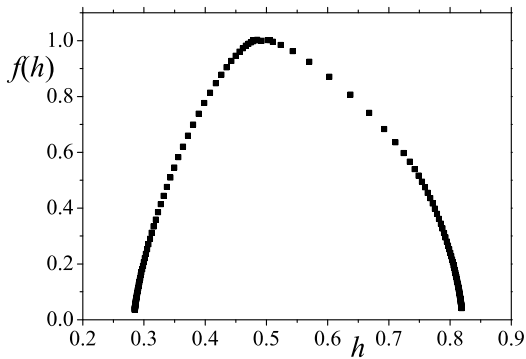}
\end{center}
\caption{Multi-fractal spectrum $f\left( h\right) $ vs. $h$ for 1 minute return
averaged over the 30 equities with $h_{\min }=0.28\pm 0.04$ and $h_{\max}=0.83\pm 0.04$.}
\label{fig-4}
\end{figure}

Let us focus now  on the multi-fractal structure of {\it return} time
series.  It has been first conjectured, and later proved, for a variety of nonextensive one-dimensional systems, that the following relation holds \cite{lyra}: 
\begin{equation}
\frac{1}{1-q_{sens}}=\frac{1}{h_{\min }}-\frac{1}{h_{\max }} \,,
\label{sens-alfa}
\end{equation}
where $h_{\min }$ and $h_{\max }$ are respectively the minimal and maximal $h$-values of the associated multifractal spectrum $f(h)$. 
In fig.~\ref{fig-4} we depict the multifractal spectrum of 1 minute traded volumes, 
obtained by the application of the MF-DFA5 method \cite{mf-dfa}; $h$ and $f\left( h\right) $ 
have been obtained from averages of the empirical data of 
30 companies. Through this analysis, we have determined $h_{\min }=0.28\pm 0.04$
and $h_{\max }=0.83\pm 0.04$. The use of Eq.~(\ref{sens-alfa}) yields $q_{sens}=0.58\pm 0.10$.
Considering that the $q$ value obtained for the return probability density function  was $q_{stat}=1.31\pm 0.02$, we verify that the dual relation
\begin{equation}
q_{stat}+q_{sens}=2
\end{equation}
is approximately satisfied within the error intervals. Taking into account the well-known fast decay of
return self-correlations, we see that the price changes for a
typical DJ30 stock may be essentially described by the $q$-triplet $\left\{
q_{sen},q_{stat},q_{rel}\right\} =\left\{ 0.58\pm 0.10,1.31\pm 0.02,1\right\} 
$.

\section{Final remarks}

In this article we have presented a nonextensive statistical mechanics
approach to the dynamics of financial markets observables, specifically the
return and the traded volume. With this approach we have been able to
present mesoscopic dynamical interpretations for the emergence of the entropic index $q$
frequently obtained by a numerical adjustment for data PDF of eqs. (\ref%
{prob-fokker}) and (\ref{prob-obt}). For the case of returns, $q$ is related
to the reaction degree of the agents on the market to fluctuations of the
observable, while for the case of traded volume it is associated to
fluctuations on the (local) average traded volume. Along with these
dynamical scenarios, and based on the multi-fractal nature of returns, we have verified 
that this quantity appears to approximatively satisfy the dual relation, $q_{stat}+q_{sens}=2$, 
previously conjectured within the emergence
of the $q$-triplet which characterises the stationary state, the sensitivity to initial
conditions, and the relaxation for nonextensive systems. The complete understanding of these connections remains ellusive. For instance, concerning relaxation
and the $q$-triplet conjecture, a new question arise for price changes. It
is well-known that the self-correlation for returns is of exponential kind, in contrast with 
the long-lasting correlations for the volatility (or returns magnitude)\ 
\cite{hes-intro}. The latter is also considered a stylised fact and it is compatible with a $q$%
-exponential form. In this way, if the efficient market hypothesis is
considered the key element in financial markets, then it makes sense to assume $q_{rel}=1$. But, if arbitrage on markets is considered as the fundamental feature instead, then 
the essential relaxation to be taken into account might be the one related to the volatility, for which $q_{rel}> 1$. Progress is clearly still needed, at both the fundamental and applied levels, in order to achieve a deep understanding of this complex system. 

\bigskip

We thank Olsen Data Services for the data provided and used herein.
Financial support from PRONEX and CNPq (Brazilian agencies), FCT/MCES
(Portuguese agency), and SI International and AFRL (USA agencies), is also acknowledged.

\end{document}